\documentclass[a4paper,fleqn]{cas-dc}

\usepackage[authoryear]{natbib}
\def\tsc#1{\csdef{#1}{\textsc{\lowercase{#1}}\xspace}}
\tsc{WGM}
\tsc{QE}
\tsc{EP}
\tsc{PMS}
\tsc{BEC}
\tsc{DE}
\usepackage{subcaption}
\usepackage{caption}
\usepackage{amsmath,amsopn}
\usepackage{physics}
\usepackage{xparse}
\usepackage{multirow}
\usepackage[graphicx]{realboxes}
\usepackage[switch]{lineno}
\usepackage{hyperref}

\begin{document}
%\linenumbers
\let\WriteBookmarks\relax
\def\floatpagepagefraction{1}
\def\textpagefraction{.001}
\shorttitle{Arrhenius activation energy and transitivity in fission-track annealing equations}
\shortauthors{M. Rufino, S. Guedes}

\title [mode = title]{Arrhenius activation energy and transitivity in fission-track annealing equations}

\author[1]{M. Rufino}[type=editor,
                        auid=000,bioid=1,
                        prefix=,
                        role=,
                        orcid=]
\ead{rufino@ifi.unicamp.br}

%\credit{Conceptualization of this study, Methodology, Software}

\address[1]{Departamento de Raios Cósmicos e Cronologia, Grupo de Cronologia, Instituto de Física ``Gleb Wataghin", Universidade Estadual de Campinas, R. Sérgio Buarque de Holanda, 777 - Cidade Universitária, Campinas - SP, 13083-859, Brazil}

\author[1]{S. Guedes}[type=editor,
                        auid=000,bioid=1,
                        prefix=,
                        role=,
                        orcid=0000-0002-7753-8584]

\cormark[1]
\ead{sguedes@ifi.unicamp.br}
\cortext[cor1]{Principal corresponding author}

\begin{abstract}
Fission-track annealing models aim to extrapolate laboratory annealing kinetics to the geological timescale for application to geological studies. Model trends empirically capture the mechanisms of track length reduction. To facilitate the interpretation of the fission-track annealing trends, a formalism, based on quantities already in use for the study of physicochemical processes, is developed and allows for the calculation of rate constants, Arrhenius activation energies, and transitivity functions for the fission-track annealing models. These quantities are then obtained for the parallel Arrhenius, parallel curvilinear, fanning Arrhenius, and fanning curvilinear models, fitted with Durango apatite data. Parallel models showed to be consistent with a single activation energy mechanism and a reaction order model of order $\approx -4$. However, the fanning curvilinear model is the one that results in better fits laboratory data and predictions in better agreement with geological evidence. Fanning models seem to describe a more complex picture, with concurrent recombination mechanisms presenting activation energies varying with time and temperature, and the reaction order model seems not to be the most appropriate. It is apparent from the transitivity analysis that the dominant mechanisms described by the fanning models are classical (not quantum) energy barrier transitions.
\end{abstract}

%\includegraphics{figs/grabs.pdf}

%\item A formalism for calculating rate constants, Arrhenius activation energies and transitivity functions is developed
%\item Results apply to any fission-track annealing model
%\item General mechanisms incorporated by each model become more apparent


 \ExplSyntaxOn 
 \cs_gset:Npn 
 \__first_footerline: { \group_begin: \small \sffamily 
 \__short_authors: \group_end: } \ExplSyntaxOff

\maketitle
\section{Introduction}
\label{Intro}

Fission-track annealing equations describe the thermal kinetics of length reduction of etched fission tracks, which are proxies for the kinetics of restoration of mineral structures damaged by fission fragments. Annealing kinetics is constrained in the laboratory through heating experiments at constant temperature ($T$) and fixed duration ($t$) and is empirically described by Arrhenius type equations, represented as isoretention ($r$) lines in a pseudo Arrhenius plot ($\ln t \times T^{-1}$). The main purpose of the annealing equations is to provide a means to extrapolate the annealing kinetics to temperatures and times characteristic of geological processes \citep{Ketcham1999, guedes2008Extrapolation}. This information allows the inference of the thermal history that a mineral sample has experienced by comparing predictions of fission-track age and length distribution with the ones measured in the sample \citep{Ketcham2019}. The reliability of the geological extrapolation from the laboratory time scale, typically up to 1000 hours, to the range of tens to hundreds of million years covered by the Fission-Track Thermochronology is a major difficulty in the application of this method \citep{Jonckheere2003Extrapolation}. Extrapolation of annealing equations is validated by comparison of model predictions with geological benchmarks, which are track lengths of samples whose thermal histories are determined by independent methods.

Extrapolation is still unsafe because annealing mechanisms are not well understood. Increasing the knowledge of the physical mechanisms underlying track annealing is essential to improve extrapolation reliability and accuracy of geological studies. Experimental studies with Small Angle X-ray Scattering (SAXS) and High Resolution Transmission Electron Microscopy (HTEM) have been carried out, revealing morphological \citep{Nadzri2015Ap, Nadzri2017Ap, Li2010Ap, Schauries2014Ap} and annealing kinetics \citep{Li2011ApvsZr, Li2012Ap} features of the unetched ion tracks in minerals at the nanometer scale. There are, however, many aspects to be clarified concerning the atomistic mechanisms.

Green and coworkers \citep{laslett1987thermal, green1988can} realized that the annealing of fission tracks is not a first-order kinetics process. The fanning Arrhenius equation proposed by them \citep{laslett1987thermal} to fit data implies in activation energy varying with the degree of length reduction. Further studies on the annealing of fission tracks in apatite and zircon suggested that the picture is still more complicated, as the fanning Curvilinear models have been shown to be more suitable for describing laboratory and geological data \citep{ketcham2007improved, guedes2013improved, Tamer2020a}. Such behavior implies that the activation energy must vary also with temperature. Physicochemical studies in condensed phase (solids and liquids) reactions have shown that this behavior is characteristic of multi-step energy activated processes depending on the reaction medium. \citet{vyazovkin2016time} presents techniques for retrieving information on the mechanisms taking place in varying activation energy systems. Aquilant and coworkers \citep{aquilanti2017kinetics, carvalho2019temperature} also tackled the problem but using the transitivity function (the inverse of the activation energy). Both approaches are complementary and help shedding light on the problem of variable activation energies. Their application to the fission track equations is not straightforward. The annealing equations must be represented in terms of reaction rates ($k$), the proper framework for calculating the activation energy.

In the next sections, a formalism to calculate reaction rates, Arrhenius activation energies, and transitivity functions from Arrhenius type fission-track annealing equations is presented.The insights on the annealing mechanisms are discussed.

\section{Method}
\label{Method}

\subsection{Annealing equations}
\label{sec:AnnealingEq}

Data sets on the annealing of fission tracks in apatite \citep{Green1986, Carlson1999data, Barbarand2003a, Tello2006}, zircon \citep{Tagami1998Zr}, titanite \citep{Jonckheere2000} and epidote \citep{Naeser1970Epidote, haack1976Epidote} have been built with the aim of studying the annealing kinetics of fission tracks. Arrhenius type annealing equations have been proposed to fit these data and make possible the extrapolation to geological timescales \citep{laslett1987thermal, LaslettGabraith1996, Guedes2005, Guedes2007, guedes2013improved, Rana2021}. Fission-track annealing equations present the general format:

\begin{equation}
    g(r)=f(t,T)
    \label{eq:FuncArr}
\end{equation}

\begin{figure}[ht!]
     \centering
     \includegraphics[width=1\linewidth]{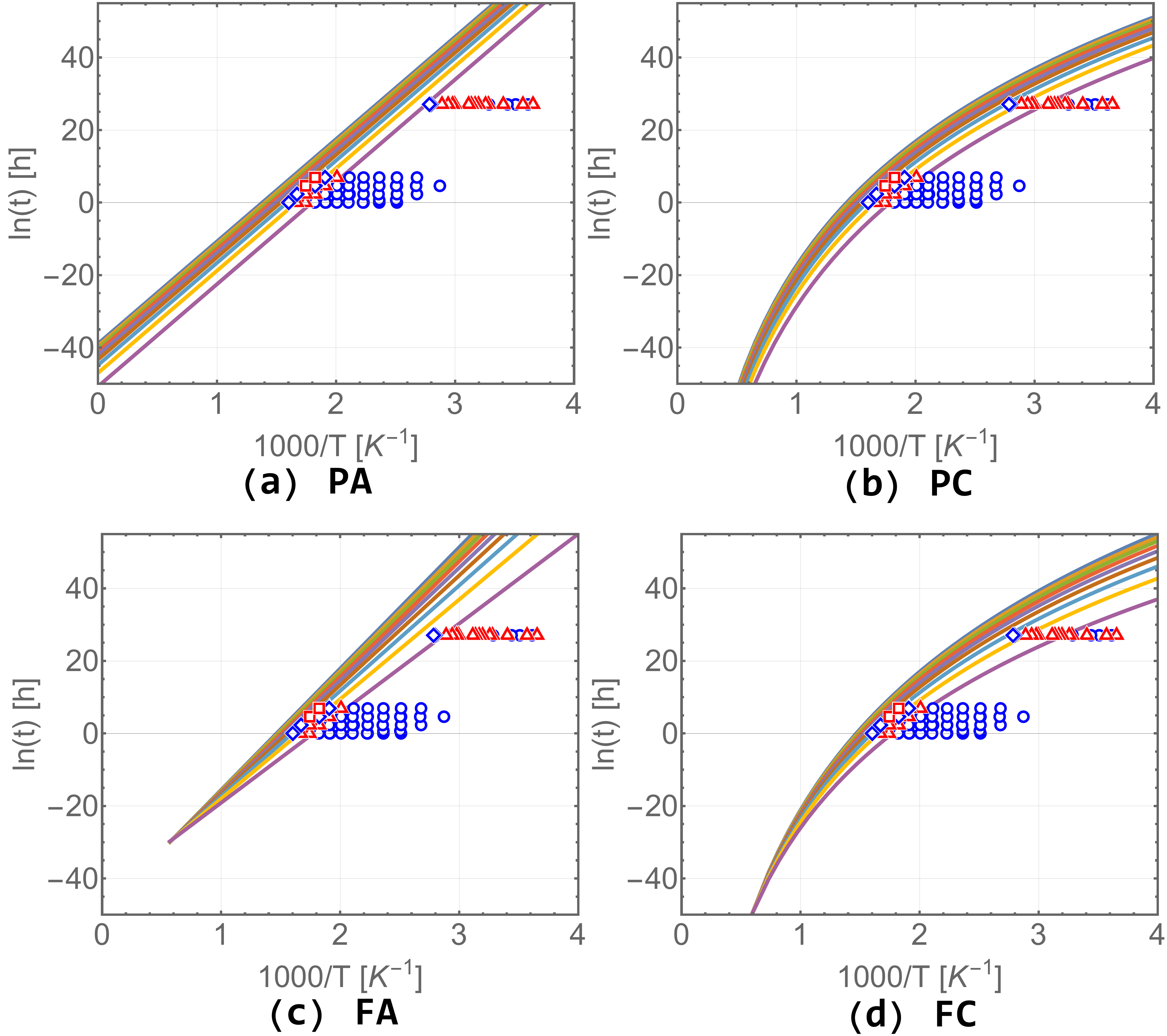}
     \centering
    \includegraphics[width=0.45\linewidth]{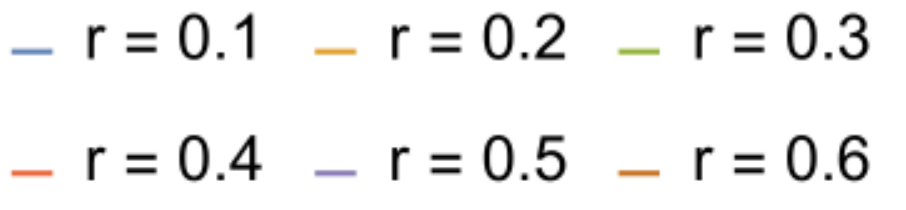}
    \includegraphics[width=0.45\linewidth]{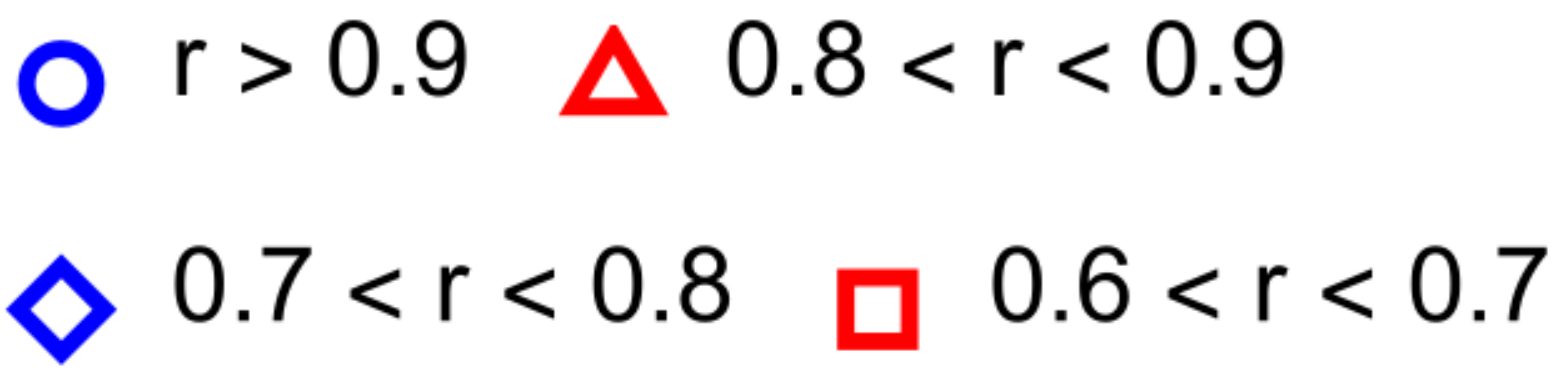}
    
    \caption{The Arrhenius pseudo-space for constant temperature experiments. Models are represented as isoretention contours. For this representation, the parameters shown in the third column of Table \ref{tab:synthesis} were used. Symbols represent laboratory Durango apatite \citep{Carlson1999data} and geological KTB apatite \citep{Wauschkuhn2015KTB} annealing data. KTB data is only included as reference for geological timescale annealing.}
    \label{fig:PAPCA pseudo arrhenius}
\end{figure}

In the above equation, $g(r)$ is a transformation of the reduced length, $r=L/L_0$. Several formats of $g(r)$ have been proposed \citep{laslett1987thermal}, being the Box-Cox transformation the most used one \citep{laslett1987thermal,ketcham2007improved}. For this work, in order to simplify the calculations and the visualization of the results (without loss of generality), the function $g(r)=\ln(1-r)$ was preferred. The function of the duration of heating ($t$) in constant temperature ($T$) annealing experiments, $f(t,T)$, defines the geometrical characteristics of isoretention contours in the Arrhenius pseudo space ($\ln t \times T^{-1}$). There are four main forms for $f(t,T)$ that are shown in Table \ref{tab:synthesis} (Eqs. \eqref{eq:PA}, \eqref{eq:PC}, \eqref{eq:FA}, and \eqref{eq:FC}). The universal gas constant ($R = 1.987204258 \times 10^{-3}$ kcal mol $^{-1}$ K$^{-1}$) is introduced in the equations in order to have activation energies given in units of kcal/mol. In addition, the argument of $\ln t$ is normalized by seconds while the argument of $\ln (1/RT)$ is normalized by mol/kcal. The parallel Arrhenius equation (PA, Eq. \eqref{eq:PA}) is represented by parallel and straight isoretention contours (Fig. \ref{fig:PAPCA pseudo arrhenius}a). The parallel curvilinear equation (PC, Eq. \eqref{eq:PC}) is represented by parallel curved contours (Fig. \ref{fig:PAPCA pseudo arrhenius}b). Fanning Arrhenius equation (FA, Eq. \eqref{eq:FA}) isoretention contours emerge from a common point ($T_0^{-1}$, $\ln t_0$), are straight and have different slopes (Fig. \ref{fig:PAPCA pseudo arrhenius}c). Fanning Curvilinear equation (FC, \eqref{eq:FC}) isoretention contours also emerge from a fanning point but are curved (Fig. \ref{fig:PAPCA pseudo arrhenius}d). For all equations, $c_0$, $c_1$, $c_2$, $c_3$, and $c_4$ are fitting parameters. For the application in this work, fitting parameters were found by fitting of c-axis projected fission track lengths ($L_{c,mod}$) of the Durango apatite annealing data set published by \citet{Carlson1999data}. This data set has been chosen because it is the same used in previous studies \citep{ketcham2007improved, Rana2021}, facilitating comparison of fittings. 

The numerical determination of parameter values was carried out through the \texttt{nlsLM} function found in the \texttt{minpack.lm} package written in \texttt{R} language. Parameter values, as well the reduced chi square ($\chi_\nu^2$) values for the fittings are shown in Table \ref{tab:synthesis}. 
%For sake of comparison, the $\chi_\nu^2$ value was calculated for the fitting of the fanning curvilinear equation (with Box-Cox transformation as $g(r)$), using the same data, presented by \citet{ketcham2007improved}. The result was $\chi_\nu^2$=2.05.

%%%%%%%%%%%%%%%%%%%%%%%%%%%
\begin{table*}[]
%\label{tab:synthesis}
\caption{Arrhenius, activation energy and transitivity equations associated with the fission-track annealing models}
\setlength\abovedisplayskip{0pt}
\setlength\belowdisplayskip{0pt}
\begin{tabular}{lcl}
\hline
FT models &  Equations    & Parameters   \\ \hline
\begin{tabular}[c]{@{}c@{}} Parallel Arrhenius \\ (PA) \end{tabular}        & \begin{tabular}[l]{@{}p{9.5cm}@{}}
\begin{equation}   
f_{PA}(t,T)= c_0 + c_1 \ln(t) + \frac{c_2}{R T}  
\tag{PA1} \label{eq:PA}  
\end{equation}
\begin{equation}
        \ln k_{ef,PA}(t,T) = \ln(\frac{c_1}{t}) - (n-1)\left[ c_0 + c_1 \ln(t) + \frac{c_2}{RT} \right]
        \tag{PA2}
        \label{eq:lnkPA}
 \end{equation}
\begin{equation}
         E_{a,PA}(t,T) = (n-1)\,c_2
         \tag{PA3}
         \label{eq:EaPA}
\end{equation}
\begin{equation}
          \gamma_{PA}(t,\beta)=\frac{1}{(n-1)c_2}
          \tag{PA4}
          \label{eq:gammaPA}
\end{equation}
              \end{tabular}
& \begin{tabular}[c]{@{}l@{}} $c_0$ = 5.631017 (dimensionless)\\ $c_1$ = 0.186520 (dimensionless)\\ $c_2$ = -10.455390 kcal/mol\\ $\chi_\nu^2$ = 2.65\end{tabular} \\ \hline
\begin{tabular}[c]{@{}c@{}} Parallel curvilinear \\ (PC) \end{tabular}       &  \begin{tabular}[l]{@{}p{9.5cm}@{}}
\begin{equation}
        f_{PC}(t,T) = c_0 + c_1 \ln(t) +c_2\ln\left(\frac{1}{R T} \right)
    \tag{PC1}
    \label{eq:PC}
\end{equation}
\begin{equation}
        \ln k_{ef,PC}(t,T) = \ln(\frac{c_1}{t}) - (n-1)\left[ c_0 + c_1 \ln(t) + c_2\ln(\frac{1}{RT}) \right]
        \tag{PC2}
        \label{eq:lnkPC}
\end{equation}
\begin{equation}
         E_{a,PC}(t,T) = (n-1)\,c_2\,R\,T
         \tag{PC3}
         \label{eq:EaPC}
\end{equation}
\begin{equation}
         \gamma_{PC}(t,\beta)=\frac{\beta}{(n-1)c_2}
         \tag{PC4}
         \label{eq:gammaPC}
\end{equation}
\end{tabular}
& \begin{tabular}[c]{@{}l@{}} $c_0$ = -4.910332 (dimensionless) \\ $c_1$ = 0.194445 (dimensionless) \\ $c_2$ = -9.610091 (dimensionless) \\ $\chi_\nu^2$ = 2.12 \end{tabular}  \\ \hline
\begin{tabular}[c]{@{}c@{}} Fanning Arrhenius \\ (FA) \end{tabular}       &  \begin{tabular}[l]{@{}p{9.5cm}@{}}
\begin{equation}
        f_{FA}(t,T) = c_0 + c_1 \frac{\ln(t)-c_2}{\frac{1}{RT}-c_3}
        \tag{FA1}
        \label{eq:FA}
\end{equation}
\begin{equation}
        \ln k_{ef,FA}(t,T) = \ln(\frac{c_1/t}{\frac{1}{RT}-c_3})-(n-1) \left[ c_0 + c_1 \frac{\ln t - c_2}{\frac{1}{RT}-c_3} \right]
        \tag{FA2}
        \label{eq:lnkFA}
\end{equation}
\begin{equation}
         E_{a,FA}(t,T) = \frac{1}{\frac{1}{RT}-c_3} \left[ 1 - (n-1)c_1 \frac{\ln(t)-c_2}{ \frac{1}{RT}-c_3} \right]
         \tag{FA3}
         \label{eq:EaFA}
\end{equation}
\begin{equation}
         \gamma_{FA}(t,\beta)= \left( \beta - c_3 \right) \left[ 1 - (n-1)c_1 \frac{\ln(t)-c_2}{ \beta-c_3} \right]^{-1}
         \tag{FA4}
         \label{eq:gammaFA}
\end{equation}
\end{tabular}
&           \begin{tabular}[c]{@{}l@{}}   $c_0$ = -8.51756 \\ (dimensionless) \\ $c_1$ = 0.12659 mol/kcal \\ $c_2$ = -20.99225 (dimensionless) \\ $c_3$ = 0.29845 mol/kcal \\ $\chi_\nu^2$ = 1.66  \end{tabular} \\ \hline
\begin{tabular}[c]{@{}c@{}} Fanning curvilinear \\ (FC) \end{tabular}       &    \begin{tabular}[l]{@{}p{9.5cm}@{}}
\begin{equation}
        f_{FC}(t,T) = c_0 + c_1 \frac{\ln(t)-c_2}{\ln\left(\frac{1}{RT}\right)-c_3}
     \tag{FC1}
    \label{eq:FC}    
\end{equation}
\begin{equation}
        \ln k_{ef,FC}(t,T) = \ln(\frac{c_1/t}{\ln(\frac{1}{RT})-c_3})-(n-1) \left[ c_0 + c_1 \frac{\ln t - c_2}{\ln(\frac{1}{RT})-c_3} \right]
        \tag{FC2}
        \label{eq:lnkFC}
\end{equation}
\begin{equation}
         E_{a,FC}(t,T) = \frac{RT}{\ln(\frac{1}{RT})-c_3} \left[ 1 - (n-1)c_1 \frac{\ln(t)-c_2}{ \ln(\frac{1}{RT})-c_3} \right]
         \tag{FC3}
         \label{eq:EaFC}
\end{equation}
\begin{equation}
          \gamma_{FC}(t,\beta)= \beta \left[\ln(\beta)-c_3 \right] \left[ 1 - (n-1)c_1 \frac{\ln(t)-c_2}{ \ln(\beta)-c_3} \right]^{-1}
          \tag{FC4}
          \label{eq:gammaFC}
\end{equation}
\end{tabular}
&  \begin{tabular}[c]{@{}l@{}}   $c_0$ = -12.80664 (dimensionless) \\ $c_1$ = 0.23109 (dimensionless) \\ $c_2$ = -41.37987 (dimensionless) \\ $c_3$ = -1.20490 (dimensionless) \\ $\chi_\nu^2$ = 1.77 \end{tabular}    \\ \hline
\end{tabular}
{\raggedright Notes: 1. For each fission-track annealing model (Eqs. \eqref{eq:PA}, \eqref{eq:PC}, \eqref{eq:FA}, \eqref{eq:FC}), the rate constants, Arrhenius activation energies and transitivity functions were obtained using $g(r) = \ln(1-r)$ and $f_r(r)=(1-r)^n$. 2. Rate constants (Eqs. \eqref{eq:lnkPA}, \eqref{eq:lnkPC}, \eqref{eq:lnkFA}, \eqref{eq:lnkFC}) were calculated after Eq. \eqref{eq:kef}. 3. Arrhenius activation energies (Eqs. \eqref{eq:EaPA}, \eqref{eq:EaPC}, \eqref{eq:EaFA}, \eqref{eq:EaFC}) were calculated using Eq. \eqref{eq:Eafr}. 4. Transitivity functions (Eqs. \eqref{eq:gammaPA}, \eqref{eq:gammaPC}, \eqref{eq:gammaFA}, \eqref{eq:gammaFC}) were obtained by applying Eq. \eqref{eq:transitivityFr}.}
\label{tab:synthesis}
\end{table*}
%%%%%%%%%%%%%%%%%%%%%%%%%%%%%%

\subsection{Isochronal Arrhenius activation energy for the fission-track system}
\label{sec:ActivationEnergy}

As discussed above, fission-track annealing models are usually represented in what is called the Arrhenius pseudo-space ($1/T$, $\ln(t)$) through the isoretention contours. This representation is particularly useful for the visualization of model extrapolation to the geological time scale. However, the visualization and interpretation of the activation energy is not straightforward. To facilitate this interpretation and comparison with documented physicochemical processes, the representation of the fission-track annealing equations in the original Arrhenius space, which relates the logarithm of the rate constant ($k$) with the inverse of temperature ($\ln k \times T^{-1}$), is required. Svante Arrhenius showed that the kinetics of chemical reactions can be described by the equation \citep{Arrhenius1889}:

\begin{equation}
	\dv{\ln k(T)}{T} = \frac{q}{RT^2}.
	\label{eq:difArr}
\end{equation}

In Eq. \eqref{eq:difArr}, $k(T)$ is the temperature dependent rate constant, $R$ is the universal gas constant and $q$ is a constant related to the change in the standard internal energy \citep[p.494]{Laidler1984Arrhenius}. Eq. \ref{eq:difArr} can be solved to:

\begin{equation}
	k(T) = A\exp\left(-E_a/RT\right).
	\label{eq:arrOriginal}
\end{equation}

In Eq. \eqref{eq:arrOriginal}, $A$ is a pre-exponential factor and $E_a$ is the Arrhenius activation energy (in units of energy per mol). In some applications the gas constant is replaced with the Boltzmann constant. In such case, the activation energy has units of energy, instead of energy per mol. For reactions obeying Eq. \eqref{eq:arrOriginal}, an Arrhenius plot ($\ln k \times 1/T$) should result in a straight line, with slope $-E_a/R$. The Arrhenius law succeed in modeling chemical reactions but, in many cases, $A$ and $E_a$ cannot be considered as constants \citep{logan1982origin, Smith2008Arrhenius}. Aware of the difficulties of interpreting the activation energies of physicochemical processes, the International Union for Pure and Applied Chemistry (IUPAC) defined the empirical quantity Arrhenius activation energy as \citep{cohen2007iupac}:

\begin{equation}\label{Ea_definicao}
    E_a = - R \dv{\ln(k)}{(1/T)}
\end{equation}

According to this definition, the Arrhenius activation energy is the slope of the Arrhenius plot at a given temperature and may depend on temperature. 

In the case of the annealing of fission-tracks, the formalism to be applied is the one used to study solid state processes. The reaction rate kinetics can be described by \citep{Vyazovkin2015Book}:

\begin{equation}
    \dv{\alpha}{t} = k(T)f_{\alpha}(\alpha)
    \label{eq:reactionRate}
\end{equation}

The function $f_{\alpha}(\alpha)$ is the reaction model. The quantity $\alpha$ in Eq. \eqref{eq:reactionRate} is the extent of conversion of a given reactant. Fission tracks can be viewed as a set of vacancies and displaced atoms. Annealing, in its turn, can be pictured as the temperature activated recombination of vacancies and displaced atoms. In this way, the conversion in fission-track annealing would be the amount of recombination events. \citet{Li2011ApvsZr} showed that the kinetics of etched track reduction follows the same trends of non-etched tracks for zircon. For apatite, tracks segment for higher degrees of annealing. This segmentation process would be more important for $r < 0.6$ \citep{Carlson1990model}, and is observed as an accelerated rate of track shortening. In spite of the differences between recombination rate and etched track shortening rate, their trends are similar. For the following analyses, the etched fission-track length ($L$) will be used as a proxy for the amount of remaining defects composing the track. The conversion and the length shortening are then linked by:

\begin{equation}
    \alpha = \frac{L_0-L}{L_0} = 1-r
    \label{eq:alpha}
\end{equation}

in which $r=L/L_0$ is the reduced track length. Substituting in Eq. \eqref{eq:reactionRate}:

\begin{equation}
    \dv{r}{t} = -k(T)f_r(r)
    \label{eq:reactionRateFT}
\end{equation}

The reaction model $f_r(r)$ describes how reaction medium influences the reaction. For the case of fission tracks, it will describe how the remaining defects, and the deformation they cause in the track region, influence the reaction rate. For instance, in a first order reaction model, the reaction (annealing) rate would be proportional to the remaining number of defects, or from a macroscopic point of view, the rate of track shortening would be proportional to the track length. \citet{green1988can} recognized that the annealing of fission tracks cannot be described by a single activation energy and that it is not a first order reaction. They showed that the fanning Arrhenius equation \eqref{eq:FA}, that implies in multiple activation energies, provides a better description of experimental data than a single activation equation \eqref{eq:PA}. In addition, it became clear that the fanning curvilinear equation \eqref{eq:FC} provides a better extrapolation to the geological time range \citep{ketcham2007improved,guedes2013improved}. Such curvature also implies that the rate constant and possibly the reaction model vary with temperature in a more complex fashion than the one implied by Eq. \eqref{eq:arrOriginal}. In similar physicochemical processes, \citet{vyazovkin2016time} suggests the use of an effective rate constant that accounts for the variation in activation energy, $k_{ef}(T)$, in Eq. \eqref{eq:arrOriginal}.
The reaction model may also change in large ranges of temperature, as it is the case of the fission-track annealing. As defect concentration diminishes and temperature increases, the conditions given for recombination may change. In this way, a more general formulation, reflecting our current lack of knowledge on the specifics of fission-track annealing is

\begin{equation}
    \dv{r}{t} = -k_{ef}(t,T)f_r(r)
    \label{eq:reactionRateFTeff}
\end{equation}

Fission-track annealing equations describe constant temperature annealing experiments carried out in different intervals of time. They are most commonly designed to produce isochronal curves, which are sets of experiments with the same duration of heating (constant $t$) and different temperatures. However, once the model parameters are found, one can retrieve the isothermal behaviour of the annealing experiments, i.e., how the degree of length reduction varies between two experiments carried out at the same temperature but in different intervals of time. From these isothermal curves, the effective rate constant can be found from Eq. \eqref{eq:reactionRateFTeff} by directly taking the derivative of the Arrhenius annealing equations. The derivative is justified in Appendix \ref{sec:AppDer}. The partial derivative is taken because the isothermal experiments (constant $T$) are being considered at this point. Firstly, both sides of Eq. \eqref{eq:FuncArr} are implicitly taken the derivative of: 

\begin{equation}
    \pdv{g(r)}{t} = \pdv{g(r)}{r}\pdv{r}{t} = \pdv{f(t,T)}{t}
\end{equation}

Then, from Eq. \eqref{eq:reactionRateFTeff}, it follows that: 

\begin{equation}
    k_{ef}(t,T) = -\frac{1}{f_r(r)}\pdv{r}{t} = -\frac{\partial_t f(t,T)}{g'(r)f_r(r)}
    \label{eq:kef}
\end{equation}

where $g'(r) \equiv \pdv{g(r)}{r}$ and $\partial_t f \equiv \pdv{f}{t}$ were defined for a more compact formulation. Assuming that the rate constant may vary during the annealing experiment, the effective rate constant given by Eq. \eqref{eq:kef} is the value of $k$ averaged over the entire experiment. In an experiment of 100 hours, for instance, the average value of $k$ during the first hour of the experiment is the same of an experiment of one our, provided the initial track length is the same in both experiments. Fig. \ref{fig:esquema_raciocinio_k} helps visualizing this concept. 
    
Applying the natural logarithm to both sides of Eq. \eqref{eq:kef}, 

\begin{equation}
    \ln k_{ef} = \ln(\partial_tf)-\ln[-g'(r)]-\ln[f_r(r)]
    \label{eq:lnkef}
\end{equation}

to obtain the Arrhenius activation energy. Thus, the variation of $\ln k_{ef}$ between two arbitrarily close temperatures on an isochronal curve is calculated, using Eq. \eqref{Ea_definicao}:

\begin{equation}
    E_a = -R \pdv{T^{-1}}\left\{ \ln \partial_tf - f(t,T) \left[ \frac{g''(r)}{[g'(r)]^2}-\frac{f'_r(r)}{f_r(r)g'(r)}\right] \right\}
    \label{eq:EaNew}
\end{equation}

\noindent in which $g''(r) \equiv \pdv{g'(r)}{r}$. Details of this derivative calculation are given in Appendix \ref{sec:AppDetails}. Eqs. \eqref{eq:lnkef} and \eqref{eq:EaNew} are pretty general, allowing for the computation of the rate constant and of the isochronal Arrhenius activation energy for any combination of $f(t,T)$, $g(r)$ and $f_r(r)$. For the analyses in the next sections, $g(r)=\ln(1-r)$ and $f_r(r)=(1-r)^n$ will be used. The latter function is a reaction-order model \citep{Vyazovkin2015Book} that has already been discussed in \citet[p.220, Eq. 8]{green1988can}. As it will be discussed later (Section \ref{sec:Results}), this reaction function produces very consistent results for the parallel models, but should be viewed as a simplification for the fanning models that allows to take into consideration more complex reaction kinetics than the ones described by first order reaction models.  

The isochronal Arrhenius activation energy becomes:

\begin{equation}
    E_a = -R \pdv{T^{-1}} \left\{  \ln \partial_tf - (n-1)f(t,T) \right\}
    \label{eq:Eafr}
\end{equation}

It is also useful to rewrite Eq. \eqref{eq:lnkef} using the chosen forms of $g(r)$ and $f_r(r)$:

\begin{equation}
    \ln k_{ef} = \ln(\partial_tf)-(n-1)\ln(1-r)
    \label{eq:lnkefFr}
\end{equation}

And, after Eq. \eqref{eq:FuncArr},

\begin{equation}
    \ln k_{ef} = \ln(\partial_t f)-(n-1)f(t,T)
    \label{eq:lnkefFtT}
\end{equation}

Before moving forward to the transitivity function, it is important to acknowledge the scope of the quantities proposed so far. Only data on constant temperature annealing experiments are available and the annealing models describe only this kind of data. For this reason, the interpretation of the rate constant and of the isochronal Arrhenius activation energy must not be extended to the variable temperature case. The rate constants are averaged over the duration of heating ($t$) and not the valuse at specific instants of time. The isochronal Arrhenius activation energy is an empirical quantity that computes variations of average chemical kinetics of annealing between experiments carried out at arbitrarily close temperatures. For these reasons, the interpretations in the next sections will concentrate on their trends for the different annealing models. For this purpose, the transitivity function is appropriate. 

\subsection{Fission-track system transitivity function}
\label{sec:transitivity}

The transitivity function ($\gamma$) is associated with the propensity of a reaction to take place and has been defined as the reciprocal of $E_a$ \citep{aquilanti2010temperature, aquilanti2017kinetics, carvalho2019temperature}:

\begin{equation}\label{eq:transi_definicao}
    \gamma\,(\beta) \equiv \frac{1}{E_a(T)}
\end{equation}

In the above equation, $\beta = 1/RT$ is the ``coldness'' \citep{muller1971coldness}, variable most appropriate for transitivity analyses. The $\gamma(\beta)$ function can be expanded in a power series and at first order becomes:

\begin{equation}
    \gamma(\beta) = \frac{1}{E_a} = \frac{1}{\varepsilon^\ddagger}+\tan{(\delta)} \beta +\order{\beta^2}
    \label{eq:linearization}
\end{equation}

The parameter $\tan{(\delta)}$ is the slope and the reciprocal of $\varepsilon^\ddagger$ is the intercept of $\gamma$ vs $1/RT$ plot. The latter parameter is the Arrhenius-Eyring energy barrier, which is the energy needed for a physicochemical process to start operating. The transitivity function regulates the transit in physicochemical transformations.

\begin{figure}[ht!]
    \centering
     \includegraphics[scale=.30]{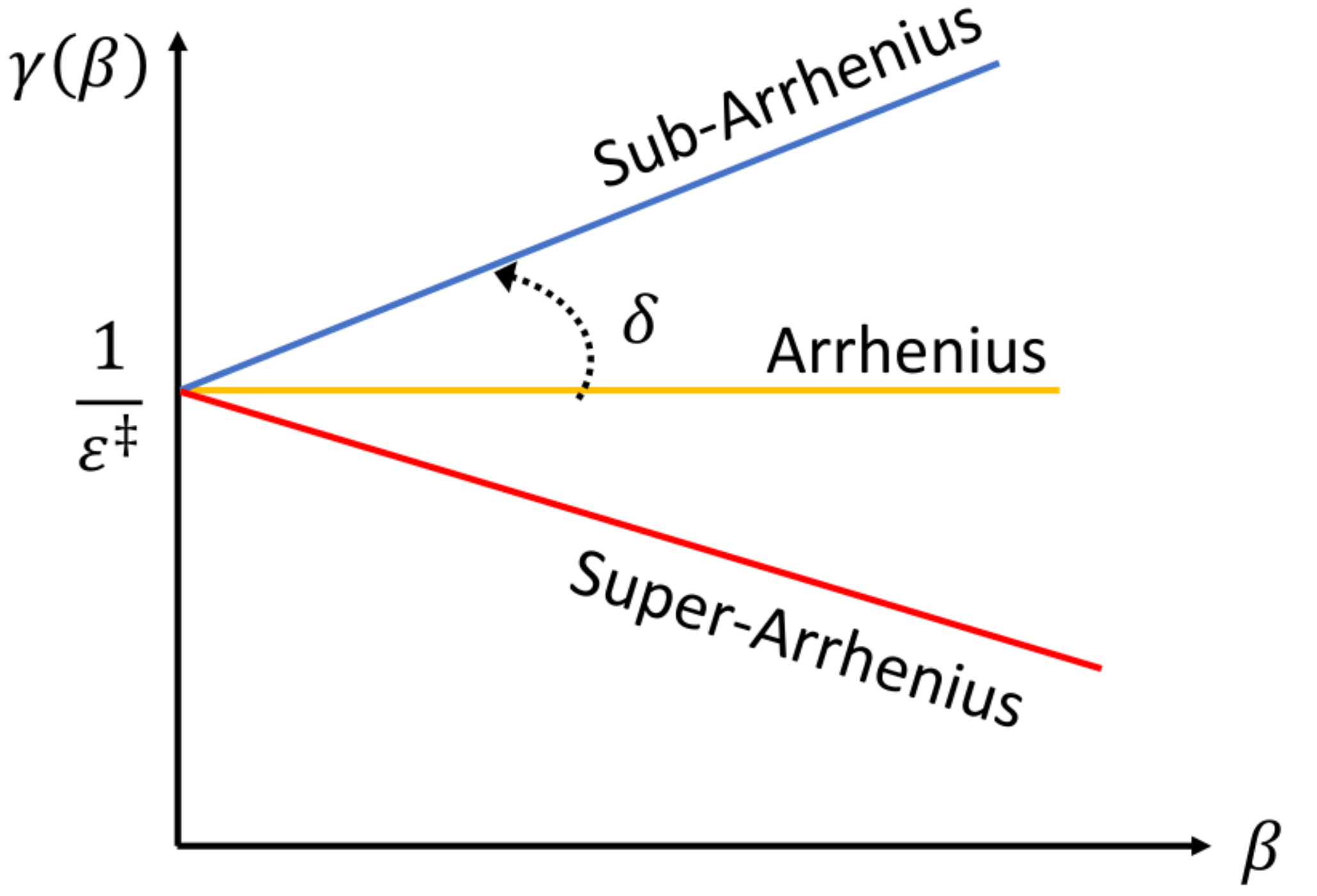}
    \caption{Schematic representation of the geometric analysis of Arrhenius deviations, its important to note that this plot shows a positive linear dependence for sub-Arrhenius and negative for super-Arrhenius in regards to the parameter $\tan{(\delta)}$ of the approximation in Eq.\ref{eq:linearization}.}
    \label{fig:transi_geom}
\end{figure}

The transitivity plot (Fig.\ref{fig:transi_geom}) highlights deviations from pure Arrhenius behaviour (constant $E_a$, $\tan \delta = 0$), thus indicating possible underlying concurrent mechanisms in the kinetics of a given physicochemical process. The method for this geometric analysis is straightforward: an horizontal line (Arrhenius line) is drawn at the point where the function $\gamma(\beta) \rightarrow 1/\varepsilon^\ddagger$. Then, the classification of the deviation emerges from the $\tan{(\delta)}$ parameter \citep{Perlmutter1976Ea, aquilanti2017kinetics,carvalho2019temperature}: 
\begin{enumerate}
    \item Sub-Arrhenius kinetics ($\tan{\delta} > 0$) will appear when quantum tunneling effects and/or complex concurrent reactions contribute to the reaction mechanism. Processes in which the introduction of $\order{\beta^2}$ terms in Eq. \eqref{eq:transi_definicao} are necessary will show an accelerated growth of $\gamma$ at lower temperatures (higher values of $\beta$).
    \item Super-Arrhenius kinetics ($\tan{\delta} < 0$) arises mainly when consecutive step mechanisms dominate complex reactions. At high values of $\beta$, the transitivity function will eventually reach a zero value, indicating the minimum temperature for the reaction to occur. Higher order terms in the $\gamma(\beta)$ expansion (Eq. \eqref{eq:transi_definicao}) will introduce a curvature in $\gamma$, shifting the minimum temperature.
\end{enumerate}

For the fission-track system, the transitivity function is the inverse of Eq.\eqref{eq:EaNew} with $\beta$ replacing $T$:

\begin{equation}
    \gamma\,(\beta) = \left\{ \left[ \frac{g''(r)}{[g'(r)]^2}-\frac{f'_r(r)}{f_r(r)g'(r)}\right] \pdv{\beta} f(t,\beta) - \pdv{\beta} \ln \partial_tf \right\}^{-1}
    \label{eq:transitivity}
\end{equation}

For the specific case of $g(r) = \ln(1-r)$ and $f_r(r) = (1-r)^n$:

\begin{equation}
    \gamma\,(\beta) = \left\{ (n-1) \pdv{\beta} f(t,\beta) - \pdv{\beta} \ln \partial_tf \right\}^{-1}
    \label{eq:transitivityFr}
\end{equation}

Through Eq. \eqref{eq:transitivityFr} and the transitivity plot (Fig. \ref{fig:transi_geom}), it is possible to investigate the Arrhenius deviation of the fission-track annealing models, gaining insights about the possible mechanisms that are operating to deviate annealing kinetics from Arrhenius.

\section{Results and Discussion}
\label{sec:Results}
In this section, the rate constants, isochronal activation energies and transitivity functions are calculated for the fission-track annealing models. Underlying mechanisms implied by each model are discussed. The equations and graphs presented in this section were developed in a script in Wolfram Mathematica Language \citep{Mathematica}, which is shared for download and visualization in \href{https://notebookarchive.org/arrhenius-activation-energy-and-transitivity-in-fission-track-annealing-equations--2022-02-c067im0/}{Notebook Archive}.

%\subsection{Fission-track annealing equations in Arrhenius space and isochronal activation energies}
%\label{sec:Arrhenius-space}

\begin{figure}[ht!]
     \centering
     \includegraphics[width=1\linewidth]{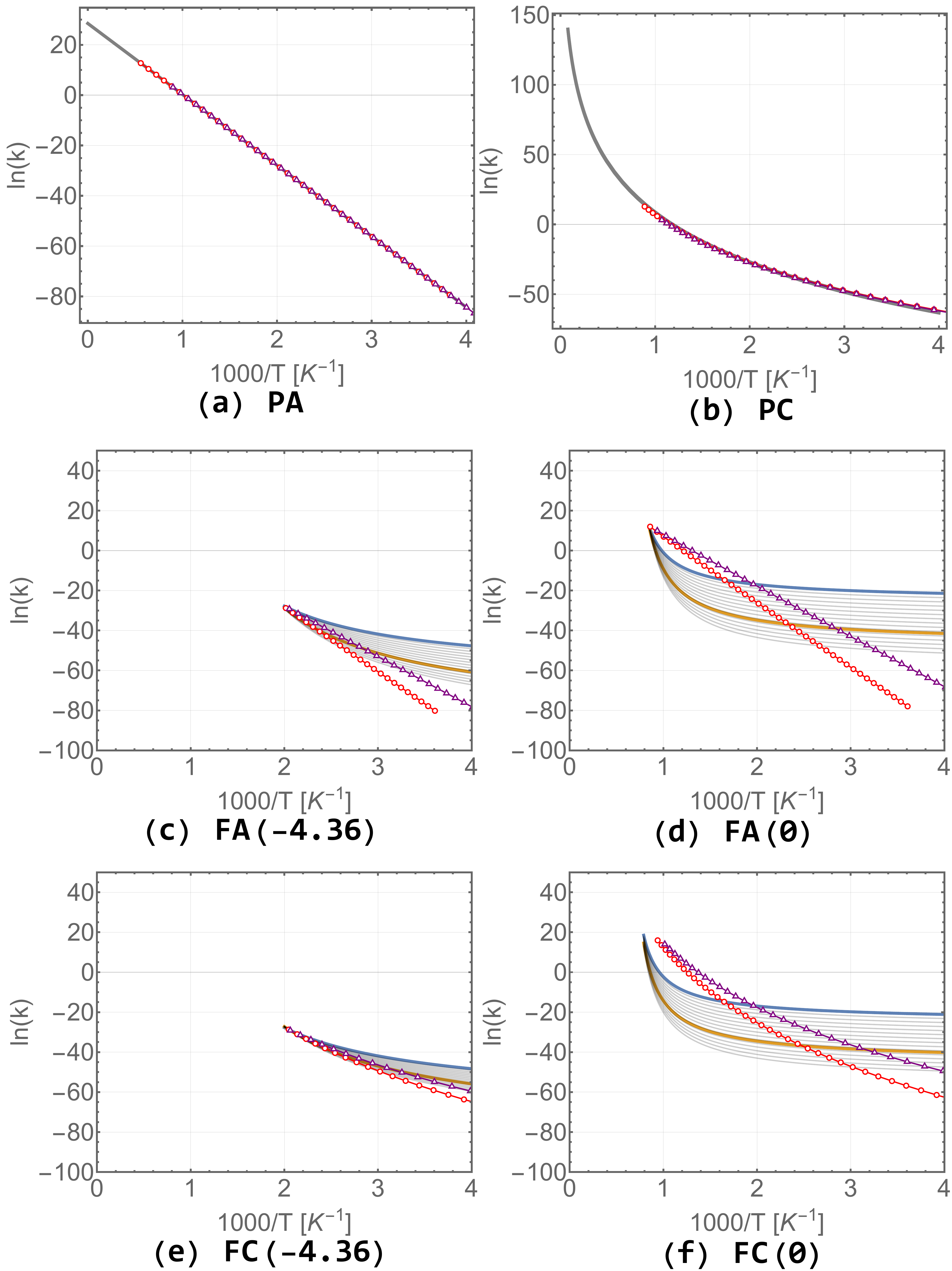}
     \centering
    \includegraphics[width=0.3\linewidth]{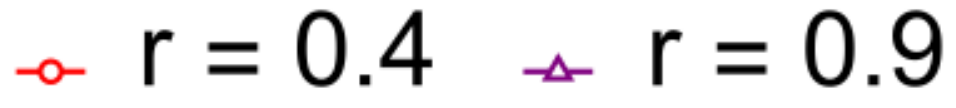}
    \includegraphics[width=0.3\linewidth]{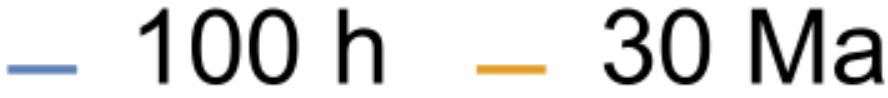}
    \caption{The Arrhenius space for constant temperature experiments. Models are represented as isochronal lines (solid) lines. The connected markers on the graph are isoretention points for $r$ = 0.4 and 0.9. Parallel models have time independent Arrhenius.}
    \label{fig:Arrhenius}
\end{figure}

Direct application of Eq. \eqref{eq:lnkefFtT} to Eqs. \eqref{eq:PA}, \eqref{eq:PC}, \eqref{eq:FA}, and \eqref{eq:FC} leads to the Arrhenius equations for the different models and are shown in Table \ref{tab:synthesis} (Eqs. \eqref{eq:lnkPA}, \eqref{eq:lnkPC}, \eqref{eq:lnkFA}, \eqref{eq:lnkFC}). Representation of $\ln k_{ef}$ for the different annealing models are shown in Fig. \ref{fig:Arrhenius}.
%Isochronal curves are shown as solid lines and represent the average rate constants for experiments carried out for a fixed time interval in different temperatures. Laboratory (100 h) and geological (30 Ma) heating conditions are represented as solid lines. 

\begin{figure}[ht!]
     \centering
     \includegraphics[width=1\linewidth]{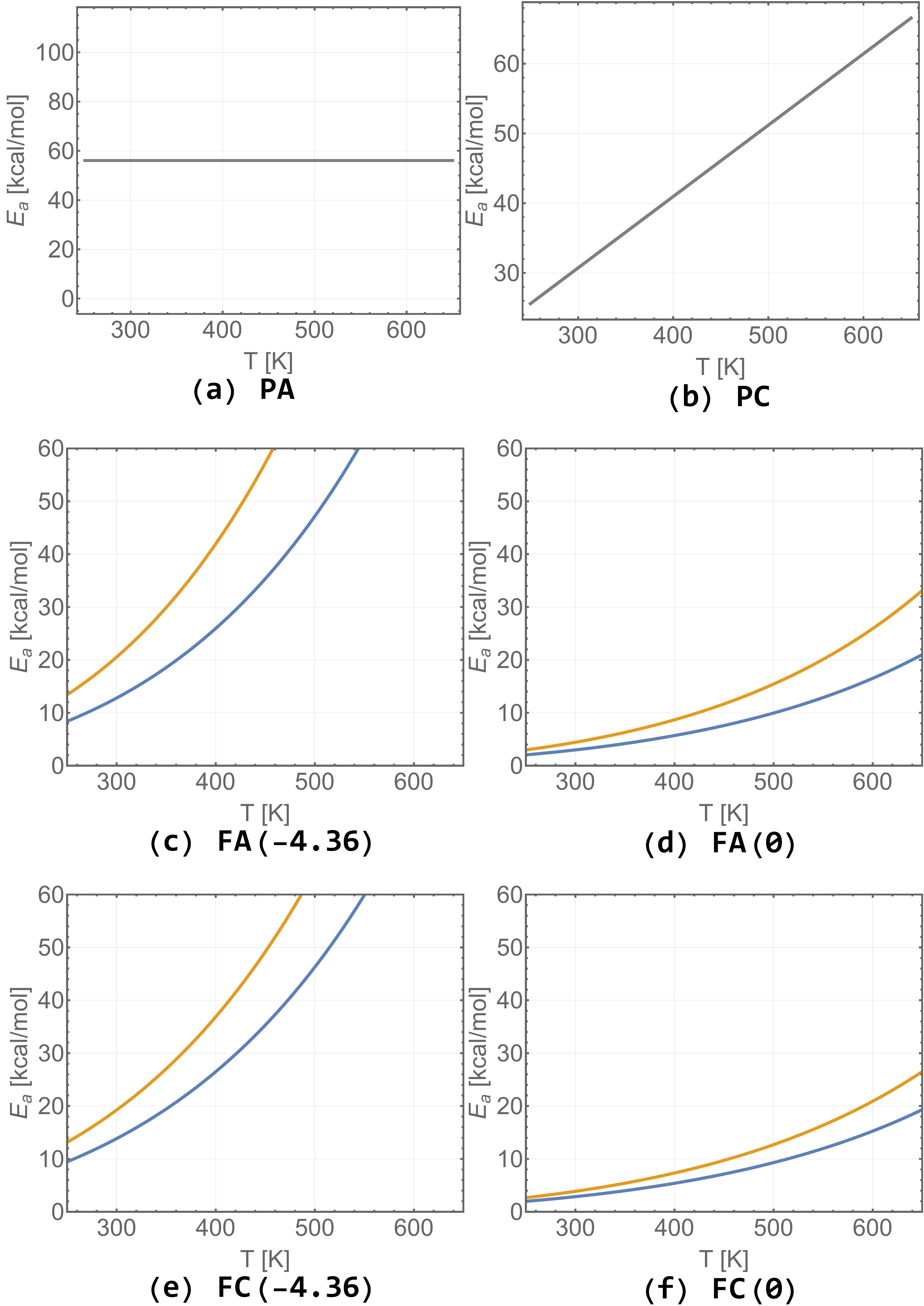}
     \centering
     \includegraphics[width=0.3\linewidth]{Ealgnd.pdf}
    \caption{Arrhenius activation energy in Fission-Track annealing models. Parallel models (PA and PC) are time independent. With the exception of PA, energies vary with temperature, revealing the possibility of an Arrhenius of multilevel processes.}
    \label{fig:Ea}
\end{figure}

The isochronal Arrhenius activation energies can be found from direct application of Eq. \eqref{eq:Eafr} resulting in Eqs. \eqref{eq:EaPA}, \eqref{eq:EaPC}, \eqref{eq:EaFA}, and \eqref{eq:EaFC}. Arrhenius activation energies for 100h and 30 Ma annealing are shown in Fig. \ref{fig:Ea}.

\begin{figure}[ht!]
     \centering
     \includegraphics[width=1\linewidth]{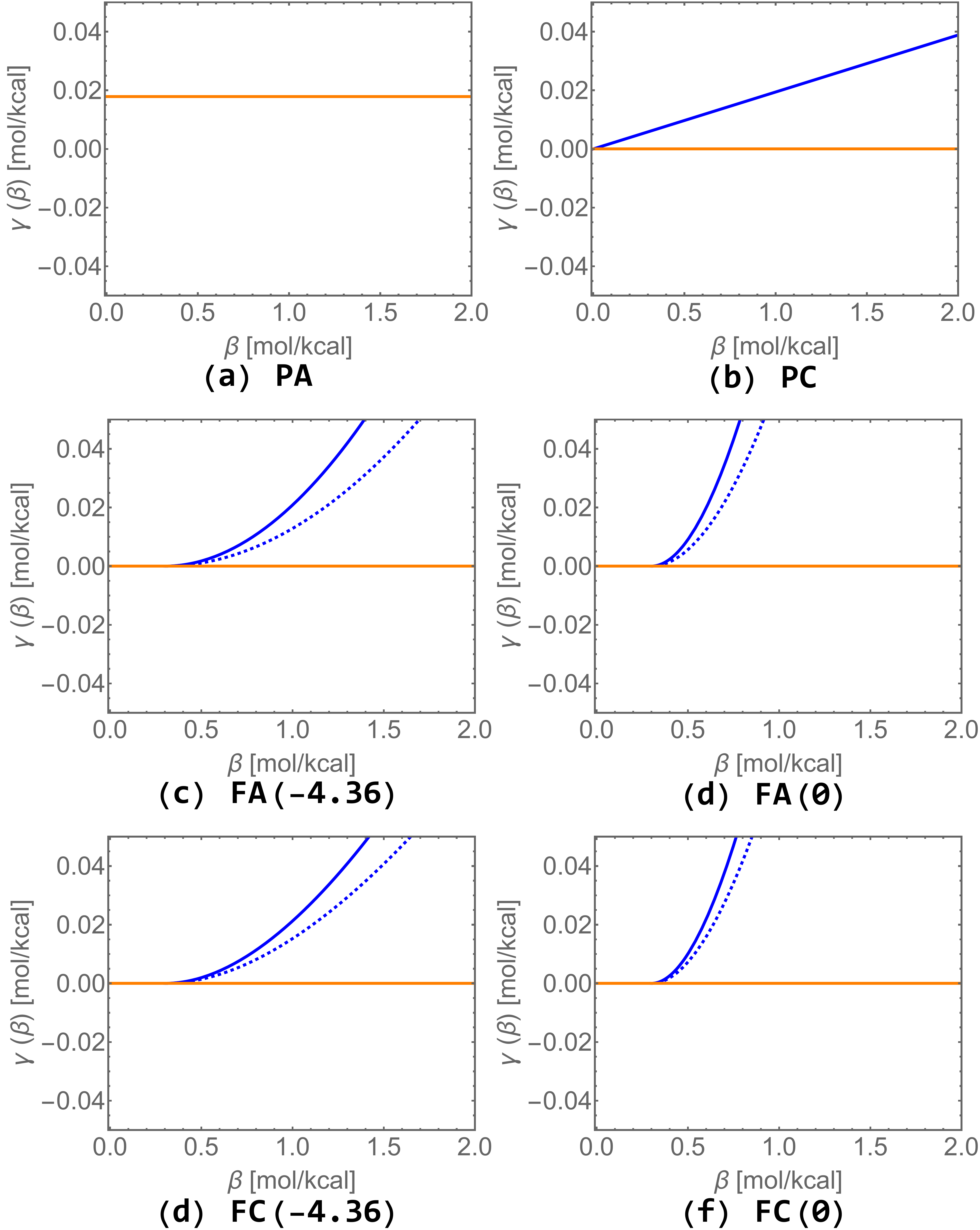}
     \centering
    
    \includegraphics[width=0.5\linewidth]{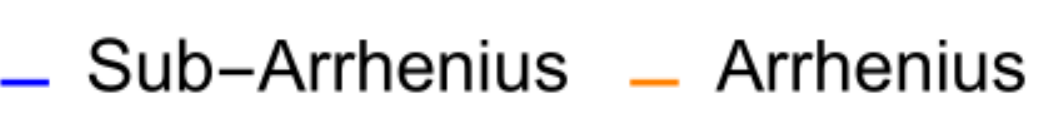}
    
    \caption{Transitivity plot in Fission-Track annealing models, the parallel Arrhenius (PA) model transitivity overlaps the Arrhenius line in the geometric method (Fig. \ref{fig:transi_geom}) revealing no transitivity. On the other hand, parallel curvilinear (PC) and the fanning (FA) e fanning curvilinear (FC) all transition from Arrhenius line to a Sub-Arrhenius regime. Moreover, the Fanning models have positive concavity in their transition, the solid line represents the experiment in laboratory time of 100 h while the dashed line represents the geological extrapolation of 30 Ma.}
    \label{fig:transitivity}
\end{figure}

The transitivity functions are obtained from Eq. \eqref{eq:transitivityFr}, resulting in Eqs. \eqref{eq:gammaPA}, \eqref{eq:gammaPC}, \eqref{eq:gammaFA}, and \eqref{eq:gammaFC} in Table \ref{tab:synthesis}. The transitivity $\gamma(\beta)$ functions are represented as isochronal contours in Fig. \ref{fig:transitivity}. From the geometric analysis (Fig. \ref{Method}), \citep{carvalho2019temperature,coutinho2021topography} it is possible to quickly identify the Arrhenius deviation at hand.

In the three figures (Figs. \ref{fig:Arrhenius}, \ref{fig:Ea}, and \ref{fig:transitivity}), results are displayed as isochronal contours of 100 h (laboratory) and 30 Ma (geological) duration of heating. For an estimation of the reaction order $(n)$, the method first proposed by \citet{green1988can} was applied. Eq. \eqref{eq:reactionRateFT} is solved by considering a single activation energy rate constant given by Eq. \eqref{eq:arrOriginal} and the reaction model $f_r = (1-r)^n$. For an annealing experiment of duration $t$, Eq. \eqref{eq:reactionRateFT} can be integrated from $t=0$ ($r=1$) to the entire duration and final track length reduction:

\begin{equation}
     \int_{1}^{r} \frac{\text{d}r}{(1-r)^n} = \int_{0}^{t} -A \exp(\frac{-E_a}{RT}) \text{d}t
     \label{eq:rateInt}
\end{equation}

The above integral can only be solved on the condition that $n \leq 0$ (considering integer reaction orders), resulting in:

\begin{equation}
    \ln(1-r) = \frac{\ln A(1-n)}{1-n} + \frac{1}{1-n} \ln t + \frac{E_a/(n-1)}{RT}
    \label{eq:rateEqSol}
\end{equation}

Remembering that $g(r) = \ln(1-r)$, the  above equation reduces to the parallel model (Eq. \eqref{eq:PA}), if
\begin{itemize}
    \item $c_0 = \ln(A(1-n))/(1-n)$
    \item $c_1 = 1/(1-n)$
    \item $c_2 = E_a/(n-1)$
\end{itemize}

For this model, the reaction order ($n$) relates to the coefficient $c_1$ by $n = 1 - 1/c_1$, resulting in $n = -4.36$. Also arises from this exercise that the activation energy for the parallel model is $(n-1)c_2$, result that is also found by applying Eq. \eqref{eq:Eafr} along with Eq. \eqref{eq:PA}. The value of $n = -4.36$ is used to draw the curves in Figs. \ref{fig:Arrhenius} (a, b, c, and e), \ref{fig:Ea} (a, b, c, and e), and \ref{fig:transitivity} (a, b, c, and e). This negative value of $n$ appears as a necessary condition in the solution of Eq. \eqref{eq:rateEqSol}, and consistently yields positive activation energies for the four annealing models. For the analyses in this section, it is initially assumed that the same reaction order model applies for the four annealing equations. 

Arrhenius isochronal contours for the parallel models collapse into a single contour, resulting in a single activation energy mechanism, independent of heating duration. The slope is constant for the parallel Arrhenius model (Fig. \ref{fig:Arrhenius}a), resulting in a single constant activation energy (Fig. \ref{fig:Ea}a, Eq. \eqref{eq:EaFA}). For the transitivity analysis, it is the standard Arrhenius behaviour (Fig. \ref{fig:transitivity}a). For the parallel curvilinear model, the slope of the Arrhenius isochronal contours vary with temperature (Fig. \ref{fig:Arrhenius}b), indicating a temperature dependent single activation energy (Fig. \ref{fig:Ea}b) annealing kinetics. The Arrhenius activation energy for this model increases linearly with temperature (Fig. \ref{fig:Ea}b, Eq. \eqref{eq:EaPC}). The isochronal contours for the parallel curvilinear model are sub-Arrhenius with no concavity (Fig. \ref{fig:transitivity}b). In this case, the deviation from the Arrhenius behaviour is due to the temperature dependence of the activation energy, which has a single value for all vacancy-defect recombination events at a given temperature.

Isoretention contours are also plotted in Fig. \ref{fig:Arrhenius} as connected circles $(r=0.4)$ and triangles $r=0.9$. For the parallel models, isoretention and isochronal contours coincide for the reaction order adopted (Fig. \ref{fig:Arrhenius}a-b), which is consistent with the interpretation that these models result from a single activation energy mechanism along with reaction order function with $n \approx -4$.

Arrhenius isochronal contours for the fanning models have different slopes, consistent with the occurrence of multi-step processes \citep{vyazovkin2016time}.They also vary with temperature and duration of heating (Fig. \ref{fig:Arrhenius}c-f). The expressions found for the Arrhenius activation energies of the fanning models (Eqs. \eqref{eq:EaFA} and \eqref{eq:EaFC}) are effective values resulting from a combination of time and temperature-dependent activation energies of concurrent processes. It is not possible at this stage to identify individual mechanisms. The fanning Arrhenius model predicts a larger variation in activation energy between experiments carried out at the same temperature than the fanning curvilinear model (compare Fig. \ref{fig:Ea}c,e and d,f).  For the fanning models, the isochronal contours present positive concavities (Fig. \ref{fig:transitivity}c-f), implying that terms of $\order{\beta^2}$ must be considered in Eq. \eqref{eq:linearization}. More complex mechanisms are then taking place, as expected from activation energy analysis. The new information provided by the transitivity plot is that the positive concavity is observed in classical (not quantum) barrier transition mechanisms during physicochemical processes \citep{carvalho2019temperature}.

Isoretention contours for the fanning models present the same characteristics as in the Arrhenius pseudo-space $(\ln t \times 1/T)$, but with the inverted sign (Fig. \ref{fig:Arrhenius}c-f). The points of intersection of isoretention and isochronal contours are the experimental conditions of temperature and duration of heating resulting in a given value of $r$. At these points, the same activation energy is expected, implying that both curves should be tangent to each other. No value of $n$ could be found that fulfilled this requirement, meaning that the reaction function $(f_r(r))$ chosen is not the most adequate to describe the behaviour captured by the fanning models. A more general format for Eq. \eqref{eq:rateInt} is found by integrating Eq. \eqref{eq:reactionRateFTeff}:

\begin{equation}
     \int_{1}^{r} \frac{\text{d}r}{f_r(r)} = \int_{0}^{t} -k_{ef}(t,T) \text{d}t
     \label{eq:rateIntGen}
\end{equation}

In this form, it is clear the interdependence between the reaction function and the effective constant rate. For the parallel models, the reaction order model is connected with single activation energy mechanisms. For the fanning models, the need for a different reaction model implies in a more complex form for the reaction rate than a single activation energy one.
This feature further separates parallel from Arrhenius model interpretations, as the latter seem to describe an even more complicated environment for track annealing. In addition to multiple concurrent thermal activated processes, the reaction medium (damaged structure) seems to play a more complicated role. Another point to be considered is that the fanning points in the Arrhenius space (Fig. \ref{fig:Arrhenius}c,e) are at much lower temperatures than expected from the fanning contours in Arrhenius pseudo-space (Fig. \ref{fig:PAPCA pseudo arrhenius}c,d). For sake of discussion, the value of $n=0$ was also used to plot the isochronal rate constants (Fig. \ref{fig:Arrhenius}d,f), activation energies (Fig. \ref{fig:Ea}d,f) and transitivity functions (Fig. \ref{fig:transitivity}d,f). This value of $n$ is still an acceptable solution for Eq. \eqref{eq:rateInt}. Note that in this limiting value, the fanning points in the Arrhenius space (Fig. \ref{fig:Arrhenius}d,f) are closer to the fanning points in the Arrhenius pseudo-space (Fig. \ref{fig:PAPCA pseudo arrhenius}c,d), which could be viewed as an indication of consistence. The reaction order of $n=0$ is the representation of an environment no influence on defect-vacancy recombination $(f_r(r)=1)$, more compatible with isolated point defects than with clustered defects forming a track. In either scenario $(n = 0 \text{ or } -4.36)$, the formats (concavities) of activation energy and transitivity functions do not change, even though the predicted values are strongly dependent on the adopted value of $n$.
Etching effects must also be taken into consideration. \citet{Jonckheere2017Etching} that, for the standard etching condition (the one applied to the data used for this work), tracks are not fully etched and that the rate of increase in track length by further etching is proportional to track length. This additional etching kinetics mechanism is superimposed to the annealing kinetics and incorporated in the effective Arrhenius activation energies.

\section{Conclusions}
\label{Conclusion}
A formalism to calculate the reaction rate $k$ (Eq. \eqref{eq:lnkef}), the activation energy $E_a$ (Eq. \eqref{eq:EaNew}), and the transitivity function (Eq. \eqref{eq:transitivity}) from annealing equations was developed, based on the literature of physicochemical processess \citep{vyazovkin2016time, carvalho2019temperature}. To apply the calculations, it was necessary to choose particular forms of the reduced length transformation $g(r)$ and of the reaction model $(f_r(r))$. The reaction order model was chosen. 

The parallel models are compatible with this reaction model along with a single activation energy of annealing. The c-axis projected data yields a reaction order of about $-4$. Altogether, this means that if annealing data is to be better represented by a parallel model, the fission-track annealing is a single activation energy process in a reaction medium (track) that can be described by a reaction order model with $n \approx -4$. The activation energy will vary with temperature if the parallel curvilinear model is chosen.

However, the fanning models have been shown to produce better fit to data \citep{laslett1987thermal, green1988can}. The fanning curvilinear model has been shown to produce the best compromise between good fits to laboratory and better agreement with geological evidence \citep{Ketcham2009,Ketcham2015, guedes2013improved}. \citet{Tamer2020a} reported the results of room temperature $(23^\circ C)$ annealing experiments with duration intervals ranging from 39 s to $\sim$ 32 years. They measured the confined track lengths of induced fission tracks in apatite samples. For c-axis projected track data in Durango apaptite, at least from the 18.3 minutes experiment, authors could fit their data, together with previous annealing data, using the fanning curvilinear model, showing that the same mechanisms captured in this model apply to the low temperature annealing range. Authors suggest that a different mechanism may be taking place in the earlier stages of annealing. Anyway, for the most part of the time and temperature ranges, the fanning curvilinear model seems to be the model that best captures annealing kinetics. This means that several activation energy mechanisms are occurring at the same time, producing the trends observed for the Arrhenius activation energy (Fig. \ref{fig:Ea}d,f) and transitivity function (Fig. \ref{fig:transitivity}d,f). Transitivity analyses also suggests that recombination occurs through classical energy barrier transitions. Fanning curvilinear activation energies have also been shown to vary during the annealing experiment, which is reflected in the different average activation energies giving rise to the the isochronas in Fig. \ref{fig:Ea}d,f.

As it is apparent from the comparison between isoretention and isochronal contours in Arrhenius plot, the reaction order model is still not the best choice for the fanning models. The expressions found under this assumption for the rate constant, activation energy and transitivity must be revisited when a more appropriate reaction model is found. For now, the reaction order model should be viewed as an approximation that allows to retrieve useful information from the trends of the fanning models.

\section*{Appendix}
\appendix
\label{Appendixes}
\section{Derivative of the fission-track Arrhenius equations}
\label{sec:AppDer}

The empirical annealing fission-track equations describe the fission-track shortening after heating at constant temperature for a given time interval. Thus, after a time interval $t_1$, the reduced track length shortens from $r=1$ to $r_1$ and the average reaction rate is:
 
\begin{equation}
    %k_1 = \frac{1-r_1}{t_1}
    \bar{\dot{r}}_1 = \frac{1-r_1}{t_1}
    \label{eq:k1}
\end{equation}
%Which in this case can be interpreted as an average annealing rate to reduce length r from its initial size 1 to $r_1$. 
Now, if the reduction is from $r = 1$ to another reduced length $r_2 (<r_1)$ in a time interval $t_2$, with the bond $t_n$ = $n \Delta t$, the average reduction rate is: 

\begin{equation}
    %k_2 = \frac{1-r_2}{t_2} = \frac{1-r_2}{2\Delta t} = \frac{k_1 + k_2}{2}
    \bar{\dot{r}}_2 = \frac{1-r_2}{t_2} = \frac{1-r_2}{2\Delta t} = \frac{\bar{\dot{r}}_1 + \bar{\dot{r}}_{1,2}}{2}
    \label{eq:k2}
\end{equation}

\noindent in which $\bar{\dot{r}}_{1,2}$ is the average reduced length rate for a reduction from $r_1$ to $r_2$ occurring in the time interval between $t_1$ and $t_2$. Then, it can be shown from Eqs. \eqref{eq:k1} and \eqref{eq:k2} that:

\begin{equation}
    \bar{\dot{r}}_{1,2} = - \frac{r_2-r_1}{\Delta t}
    \label{eq:k12}
\end{equation}

\begin{figure}[ht!]
    \centering
     \includegraphics[scale=.30]{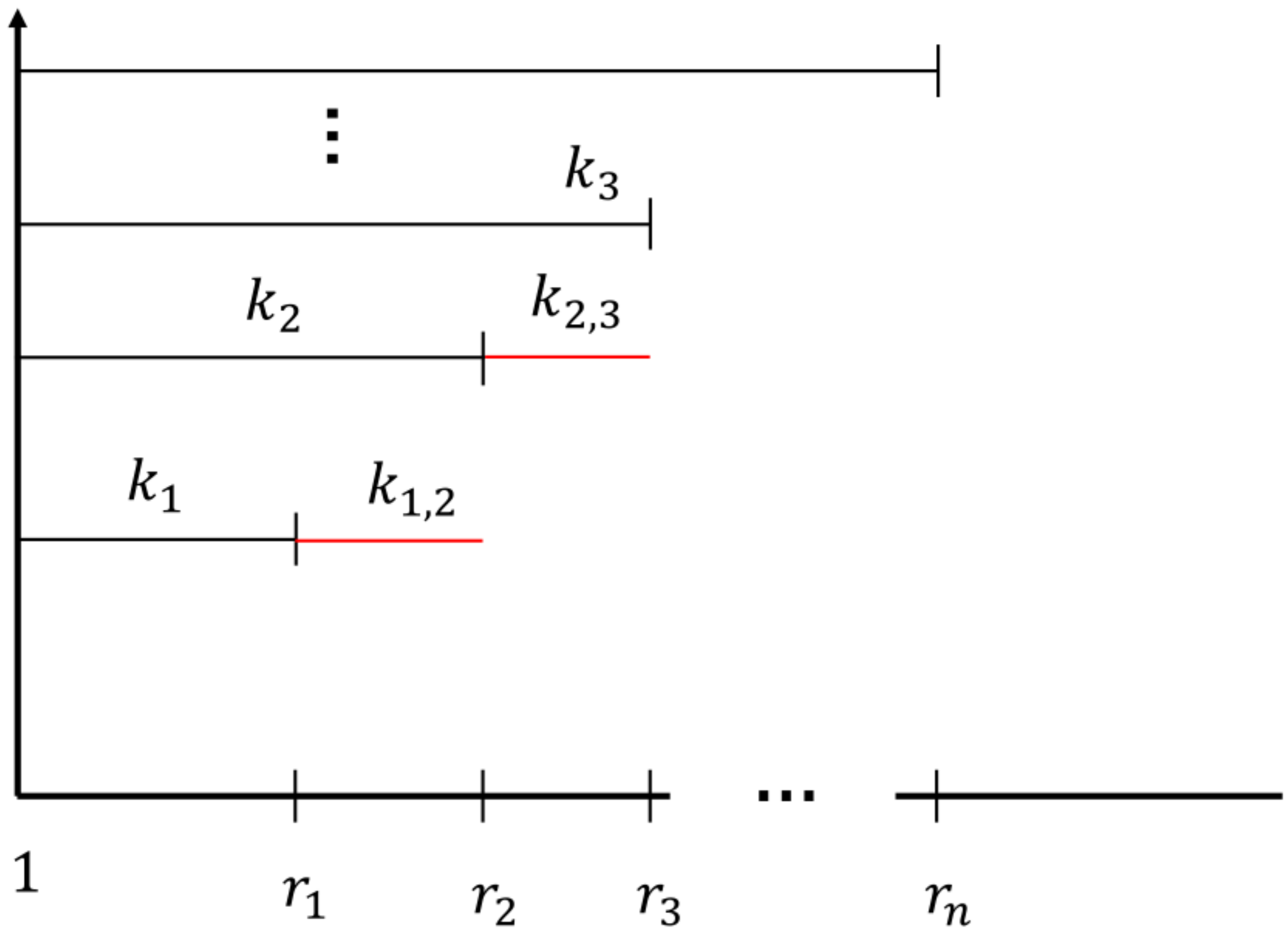}
    \caption{Representation of the mathematical induction scheme for obtaining the reaction rate k.}
    \label{fig:esquema_raciocinio_k}
\end{figure}

By mathematical induction (Fig.\ref{fig:esquema_raciocinio_k}) it is possible to generalize the result in (\ref{eq:k12}) from $r_{n-1}$ to $r_n$, thus obtaining a reduction rate $\bar{\dot{r}}_{ n-1,n}$: 

\begin{equation}
    \bar{\dot{r}}_{n-1,n} = - \frac{r_n - r_{n-1}}{\Delta t} = -\frac{\Delta r}{\Delta t}
\end{equation}

Finally, by approximating $\Delta t$ to a short enough interval, this annealing rate can be defined as a local annealing rate: 

\begin{equation}
    \dot{r} = -\dv{r}{t}
    \label{eq:instantReactionRate}
\end{equation}

Eq. \eqref{eq:instantReactionRate} assures that the annealing rate appearing in Eq. \eqref{eq:reactionRateFTeff} can be calculated from the annealing Eqs. \eqref{eq:PA}, \eqref{eq:PC}, \eqref{eq:FA}, and \eqref{eq:FC}. In the application of Eq. \eqref{eq:instantReactionRate}, the temperature must be kept constant as this is the initial hypothesis of the above deduction and the equation can be written as:

\begin{equation}
    \dot{r} = -\pdv{r}{t}
    \label{eq:instantReactionRatePartial}
\end{equation}

\section{Details of the Arrhenius activation energy calculation}
\label{sec:AppDetails}

In this appendix, the details for the derivative of $\ln k_{ef}$ with relation to temperature, $T$ are given. Firstly, as the derivative is to be taken over an isochronal curve (constant duration of annealing) the partial derivative of Eq. \eqref{eq:lnkef} is to be calculated. As $r$ is also a function of $T$, the chain rule must be applied:

\begin{equation}\label{lnk1}
    \pdv{\ln(k)}{T^{-1}} = -\frac{g''(r)}{g'(r)}\pdv{r}{T^{-1}} + \pdv{\ln[\partial_tf(t,T)]}{T^{-1}}
\end{equation}

in which $g''(r) \equiv \pdv{g'(r)}{r}$. It is convenient to remove $\partial r/\partial T^{-1}$ from the previous equation. For this, the following relation (also from the chain rule) is used: 

\begin{equation}
    \pdv{g(r)}{T^{-1}} = \frac{g'(r)}{\pdv*{r}{T^{-1}}} = \pdv{f(t,T)}{T^{-1}}
\end{equation}

And, 

\begin{equation}
    \pdv{r}{T^{-1}} = \pdv{f(t,T)}{T^{-1}}\frac{1}{g'(r)}
\end{equation}

Then the Eq.\eqref{lnk1} becomes: 

\begin{equation}
    \pdv{\ln(k)}{T^{-1}} = -\frac{g''(r)}{[g'(r)]^2}\pdv{f(t,T)}{T^{-1}} + \pdv{\ln[\partial_tf(t,T]}{T^{-1}}
\end{equation}

Finally, from the definition, Eq.\eqref{Ea_definicao}, the Arrhenius activation energy becomes: 

\begin{equation}\label{Ea_STF}
    E_a = R \frac{g''(r)}{[g'(r)]^2}\pdv{f(t,T)}{T^{-1}} - R \pdv{\ln[\partial_tf(t,T)]}{T^{-1}}
\end{equation}

\section*{Acknowledgements}
This work has been funded grant 308192/2019-2 by the National Council for Scientific and Technological Development (Brazil).

\bibliographystyle{cas-model2-names}

\bibliography{cas-refs}

\end{document}